\documentstyle[12pt,aasms4]{article}
\begin{document}

\title{Accretion Disks Around Class 0 Protostars: \break
The Case of VLA~1623}

\author{Ralph E. Pudritz\altaffilmark{1}, Christine D. Wilson\altaffilmark{1}}
\author{John E. Carlstrom\altaffilmark{2}, Oliver P. Lay\altaffilmark{3}}
\author{Richard E. Hills\altaffilmark{4}, and Derek Ward-Thompson\altaffilmark{5}}

\bigskip
\centerline{\it Submitted to the Astrophysical Journal Letters, 17 May 1996}  

\centerline {\it Revised July 19, 1996}

\altaffiltext{1}{Department of Physics and Astronomy, McMaster University,
Hamilton, Ontario L8S 4M1 Canada} 
\altaffiltext{2}{Department of Astronomy and Astrophysics, University
of Chicago, Chicago, IL 60637}  
\altaffiltext{3}{Division of Physics, Mathematics, and Astronomy,
California Institute of Technology, Pasadena CA 91125 U.S.A.} 
\altaffiltext{4}{Mullard Radio Astronomy Observatory, Madingley
Road, Cambridge CB3 0HE, UK}
\altaffiltext{5}{Royal Observatory, Blackford Hill, Edinburgh,
EH9 3HJ, UK}
\begin{abstract}

Continuum emission at 220 and 355 GHz 
from the prototype Class 0 source VLA 1623 has been detected
using the JCMT-CSO interferometer. 
Gaussian fits to the
data place an upper limit of 70 AU on the half-width half-maximum
radius of the emission, which implies an upper limit of 
$\sim 175$ AU for the cutoff radius 
of the circumstellar disk in the system.  
In the context of existing 
collapse models, this disk could be magnetically supported 
on the largest scales and have an age of $ \sim 6 \times 10^4$~yr,
consistent with previous suggestions that Class 0 sources are quite young.
The innermost region of the disk  within
$\sim 6$ AU is likely to be in centrifugal support, which 
 is likely large enough to 
provide a drive for the outflow according to current
theoretical models. Alternatively, if 
175 AU corresponds to the
centrifugal radius of the disk, the age of the system is $\sim 2\times 
10^5$ yr, closer to age estimates for Class I sources. 

\end{abstract}


\section{Introduction}

Submillimeter observations of deeply embedded star-forming 
regions have identified a small number of strong sources  
that have virtually no emission at wavelengths below 10~$\mu$m   
and whose spectral energy distributions are
characterized by single
blackbodies at $T \simeq 15-30 $ K.
These cold ``Class 0'' sources are good     
candidates for protostars whose circumstellar
envelope masses exceed that of  their central protostars 
(\markcite{AWB}Andr\'e, Ward-Thompson, \& Barsony 1993, 
\markcite{AM94}Andr\'e
\& Montmerle 1994, \markcite{A95}Andr\'e 1995), in which case   
their ages
may be as low as $2 \times 10^4$ yrs (\markcite{B94}Barsony 1994).  
Sources in this class are also defined  
by the presence of energetic bipolar outflows.

There is now good evidence that a large fraction of young stellar objects 
have dense, dusty accretion disks
(eg. \markcite{Beckwith}Beckwith et al. 1990, \markcite{Lada}Lada 1991).   
This is in accord with standard collapse models
of star formation (eg. \markcite{Terebey}Terebey, Shu, \& Cassen 1984), 
which predict that systems older than
$10^5$ years have centrifugally supported disks extending
to 100~AU.  
Disks around the sources L1551 IRS5 and HL Tau have
been resolved with semi-major axes of  
80 and 60 AU, respectively 
(\markcite{Lay94}Lay et al. 1994, \markcite{Mundy}Mundy et al. 1996).    
Even larger magnetically supported ``pseudo'' disks 
extending out to 1000 AU are predicted for some magneto-hydrodynamic
collapse models (\markcite {GS}Galli \& Shu 1993, \markcite{Fiedler}
Fiedler \& Mouschovias 1993).

Most current models of outflow posit the existence
of disks as an essential element of the engine
(eg. \markcite{Pudritz}Pudritz, Pelletier, \& Gomez de Castro 1991, 
\markcite{Konigl}Konigl \& Ruden 1993,  \markcite{Shu}Shu et al. 1993).   
If the Class 0 sources
are young, then the Terebey et al. (1984) models suggest that
a disk with radius of 100 AU at $2 \times 10^5$ yrs has  a
centrifugal radius of only 
$r_c \propto t^3 \simeq 0.8$ AU at $2 \times 10^4$ yrs.
Similarly a larger magnetically supported
outer region of $r_B \propto t^{7/3} \simeq 20$~AU 
(\markcite{GS}Galli \& Shu 1993) should
be present at these early times if 1000 AU structures exist
at $2 \times 10^5$ yrs. Such 
small, centrifugally supported regions would still be able to drive outflows,
while the larger magnetic structure could  
play an important role in star formation itself.
This {\it Letter} describes the results of our search for an accretion disk 
in the prototype Class 0 source, VLA 1623, using the JCMT-CSO submillimeter
interferometer.

\section{Observations and Analysis}

Observations of VLA 1623 were obtained with the JCMT-CSO interferometer
on 7 April 1995 at a frequency of 356 GHz
with 500 MHz bandwidth and 9 April 1995 at 353 GHz with 1 GHz
bandwidth. The projected length of
the baseline ranged from 140 to 190 k$\lambda$
and observations were obtained over a period of two to three
hours each night. The quasar NRAO 530 was used
to monitor the gain of the instrument and the absolute uncertainty
in the flux calibration is 20\%. The average gain of the instrument was
70 Jy K$^{-1}$ and the system temperature was
800 K. The data were obtained with 10 s integrations and
vector-averaged over 100 s intervals. 
Since  the phase response of the interferometer over longer periods of
time is not yet known, only the amplitude of
the visibility function was used in the analysis.
Observations were also obtained at a frequency of 220 GHz with
1 GHz bandwidth on 10 April 1995
over a period of 7 hours. The projected length of the baseline ranged from
60 to 120~k$\lambda$ and the system temperature was 400 K. 

The visibility flux of VLA 1623 shows no dependence on the projected
baseline length (Figure~\ref{fig-1}a,b) and thus the source is unresolved. 
We measured both the
flux and an upper limit to the full-width half-maximum 
radius of the source using the techniques
outlined in \markcite{Lay94}Lay et al. (1994). 
Briefly, we fit a two-dimensional
gaussian to the data using the Rice distribution 
to describe the noise.  The position angle was fixed at $35^o$ to
orient the disk perpendicular to the large-scale outflow
(\markcite{andre90}Andr\'e et al. 1990).   Fifty  
simulated data sets were then generated by 
adding artificial noise to the best fitting model. The best fit 
parameters (flux and major-axis radius) derived from these simulated 
data sets are shown in Figure~\ref{fig-1}d, from which 90\% confidence
limits can be determined. 
At 355 GHz, the flux is $0.5^{+0.4}_{-0.1}$ Jy
and the major-axis radius is $R < 70 $ AU 
(for a distance to
$\rho$ Oph of 160 pc).   Fitting the data with
a circularly symmetric gaussian produces a somewhat smaller
upper limit to the radius of 45 AU. 
A more physically valid model would be a power-law disk (see \S 3).
The large uncertainty in the flux at
355 GHz is due to the rather limited baseline coverage obtained
for this low declination source. As a result, the model fits
are unable to distinguish between a weaker compact source 
and a stronger, more extended  source 
for which some of the flux is resolved out even on the
shortest baselines at 355 GHz
(Figure~\ref{fig-1}).
 At 220 GHz, the flux is
$0.12_{-0.02}^{+0.08}$ Jy and the radius is $R < 150 $ AU.

We calculated the expected contribution to the visibility flux from the
extended envelope as in Lay et al. (1994).
We modeled the envelope of VLA 1623 as an optically thin sphere
with $T \propto r^{-0.5}$ and $\rho \propto r^{-1.5}$.
The flux scale was calibrated
using single dish fluxes from \markcite{awb}Andr\'e et al. (1993). 
We also assumed that the envelope stops where the disk begins
and so adopted an inner cutoff radius for the envelope of 70 AU. 
Under these conditions, the envelope for 
the 70 AU model contributes negligible flux at 355 GHz
and only 0.03 Jy at 220 GHz. 
Thus, we attribute most of the observed flux to the 
presence of an unresolved disk in VLA 1623.

\section{Fitting Accretion Disk Models}

Our observations provide useful constraints on
plausible accretion disk models for the origin of the
submillimeter emission.  We adopt the approach of 
Beckwith et al. (1990) and assume a geometrically thin
disk with power law
scalings with radius of the disk temperature
and surface density, $T(r) = T_o (r/r_o)^{-q}$ and
 $\Sigma (r) = \Sigma_o (r/r_o)^{-p}$, 
where $T_o$ and
$\Sigma_o$ are  the dust temperature and mass surface density
at the inner radius  $r_o$
of the dust distribution.  The optical depth is
$\tau = \Sigma \kappa_{\nu} / cos \theta$, where $\theta $ is
the inclination of the disk axis to the line of sight and 
$\kappa_{\nu} = \kappa_o
(\nu/\nu_o)^{\beta}$ is the specific dust opacity.    
The flux emitted by the disk is then 
$F_{\nu} = (2 \pi cos \theta/ D^2) 
\int_{r_o}^{R_D} B_{\nu}(T)(1 - e^{-\tau}) r dr$
where $B_{\nu}(T)$ is the Planck function, $D$ is the distance to
the source, and $R_D$ is the outer disk cutoff radius. 
Power-law fits to disks around HL Tau and L1551 IRS5 produce cutoff radii
up to a factor of 2.5 larger than radii from gaussian fits 
(\markcite{Lay94}Lay et al. 1994,
\markcite{Mundy}Mundy et al. 1996). However, since
VLA 1623 is unresolved, there is no structure in the visibility data to
constrain a power law and so such a fit was not performed on this data set.

While $ q = 0.75$ for purely viscous accretion disk models, 
more general calculations which incorporate the effects of reprocessing
and disk flaring typically find $ 0.5 \le q \le 0.75$. 
Noting that various model fits have  $ 0.5 \le p \le 1.5 $
(eg. \markcite{Thamm}Thamm, Steinacker, \& Henning, 1994), 
we adopt ranges of  $0.25 \le q \le 0.75$ and 
 $-0.25 \le 2-p-q \le 1.25$.  
The maximum dust temperature must be less than
its sublimation temperature, $T_{o,max}\sim 2000$ K. 
The inner edge of the 
disk cannot be inside the stellar photosphere, which
for low mass stars gives $r_o \ge R_* \simeq 2$ R$_{\odot}= 0.01$ AU.
We take the inclination angle of the disk to be  
that of the outflow from VLA 1623, 
$\theta = 80^o\pm5^o$ or almost edge-on 
(\markcite{andre90}Andr\'e et al. 1990).   
The spectrum
$F_{\nu} \propto \nu^{\alpha}$ determined  
from our measured fluxes at 355 and 220 GHz has $ \alpha \simeq 
3.0^{+1.6}_{-2.2}$, where 
optically thin models have  $\alpha = 2 + \beta $ and
optically thick models have $\alpha = 2$. Recent 
calculations show that $ \beta \simeq 1.5 \pm 0.5$ 
at wavelengths from 650 $\mu$m to 2.7 mm for submicron
to millimeter sized particles and
 a wide variety of densities and temperatures (\markcite{Poll}Pollack 
et al. 1994). 
Finally, the parameter space characterizing the emission is also constrained  
by our observed flux, our upper limit to the cutoff radius of
$\sim 2.5 \times 70 = 175$ AU, and
 by the $\sim 1$ L$_\odot$ bolometric luminosity of the entire
source (\markcite{awb}Andr\'e et al. 1993).

It is obvious that 
this accretion disk model covers a large parameter space
that is relatively poorly constrained.  
The issue of the uniqueness of model fits has been addressed 
by \markcite{Thamm}Thamm et al. (1994) 
and \markcite{Bouvier}Bouvier \& Bertout
(1992), who find ``global ambiguity'' in acceptable fits to models 
of T Tauri disks based on spectral energy distributions 
that incorporate far more infrared data than is
available for VLA1623.
However, for the limiting cases of completely optically
thick or thin truncated disks in the Rayleigh-Jeans
limit,  the cutoff radius can be related to
the observed flux $F_{\nu}$ by
$$ R_D / r_o = [(\gamma \bar L_{\nu}/ x cos \theta ) + 1]^{1/\gamma}
\eqno(1)
$$ 
where $\bar L_{\nu} = D^2 F_{\nu}/ 2 \pi (2 k \nu^2/c^2)$ can
be computed from the observations.
 For the optically thick case, the parameters 
 are $\gamma = 2 - q$ and 
$ x = T_o r_o^2$, while  for the optically 
thin case, $\gamma = 2 - p - q \ne 0$ and $x = \tau_o T_o r_o^2$, where
$\tau_o \equiv \kappa_\nu \Sigma_o / cos \theta $.   

Equation (1) shows that we can simplify our analysis of the model
by reducing the six free parameters ($\kappa_\nu$, $q$, $p$, $T_o$, $r_o$,
$\Sigma_o$) to essentially the two parameters  $\gamma$ and $x$ (with
$R_D$ measured in units of $r_o$).
Then  a  disk model characterized by a given index $\gamma$  
produces a unique curve in the $R_D/r_o - x$ plane. 
These curves are plotted in Figure~\ref{fig-2} for  VLA 1623, 
for which $F_\nu = 0.5$ Jy gives $\bar L_{355} = 2.25 \times 10^4 $~K~AU$^2$.
In addition, our observed flux and upper limit to the cutoff radius
restrict possible disk models to lie within the region
demarcated by the three dashed lines in Figure~\ref{fig-2}.  
Boundary line
(A) is the $ \gamma = 2$ model (i.e. $q= 0$ or $p + q = 0$ for
optically thick or thin cases), which demarcates a model disk 
with uniform surface brightness and therefore 
the smallest cutoff radius $R_D$ for the emitted flux.  Line (B)  is
our upper limit
to the cutoff radius $R_D \le 175 $ AU coupled with the
condition that $r_o \ge 0.01 $~AU, while line  
(C) is the upper limit to $R_D/r_o$ for a given $x$. Line (C) 
arises from the fact that $T_o$ must be less than $T_{o,max}$, so that
$r_o$ must be greater than $(x/T_{o,max})^{1/2}$, and thus
$R_D \le 175$ AU implies 
$R_D/r_o \le 175 (T_{o,max}/x)^{1/2} $.   
Figure~\ref{fig-2} shows that while optically thick models with 
values of $q \simeq 0.25-0.5$
fill the largest portion of the permitted parameter space, 
models with $q$ up to 0.75 are also acceptable. 
 Optically thin models fill a much 
smaller portion of parameter space, with
low values of $2-p-q$ being less likely. 
In general,  
the optically thin 
solutions have relatively large values for the inner radius,
which suggests that they
may also have a small optically thick region that contributes
relatively little to the total flux of the source.

The observed bolometric luminosity of the envelope+disk system
 also provides a constraint
on the cutoff radius. For $q \ne 0.5$, the luminosity of
an optically thick disk is 
given by $ L_{bol} = 4\pi\sigma T_o^4 r_o^2 ((R_D/r_o)^{2-4q} - 1)/(2-4q)$,
while for $q=0.5$, 
$ L_{bol} = 4\pi\sigma T_o^4 r_o^2 ln(R_D/r_o)$.
For $q < 0.5$, this expression can be combined with equation (1) to give
$$R_D = 47.7 \hskip2pt{\rm AU} {{(2-q)^{2/3}} \over {(2-4q)^{1/6}}} 
\Big({L_{bol} \over 1 \hskip2pt{\rm L_\odot}}\Big)^{-1/6} \Big({{L_{355} 
/ 2.25 \times 10^4 \hskip2pt{\rm K \hskip2pt AU^2}} \over
{cos\theta / cos 80}}\Big)^{2/3} \eqno(2)$$
In addition, equation (1) can be rewritten as $T_o r_o^q \propto R_D^{q-2}$,
which allows the straightforward calculation of $T_o r_o^q$ for a given
$R_D$ obtained from equation (2).
For $q=0$ and $L_{bol} < 1$ L$_\odot$, 
the cutoff radius is $R_D > 68$ AU and the inner disk temperature 
is $T_o < 57$ K, while for $q=0.25$ and $r_o = 0.01-10$ AU,
the radius is $>$69 AU and the temperature is $T_o < 430-80$ K. 
For $q \ge 0.5$ the expression for $R_D$ is different from equation (2)
and depends on the value of $r_o$.
For $q=0.5$ and $r_o = 0.01-1$ AU, the cutoff radius is $>$90-85 AU, 
while for $q=0.75$ the radius is $>$230-120 AU. 
A smaller but warmer optically
thin disk could probably also produce the observed 355 GHz flux and
bolometric luminosity, but the numerical models required to
derive the exact cutoff radius in this case are beyond the scope of this paper.

Thus both the spectral index and the models summarized in Figure~\ref{fig-2} 
are consistent with either
 optically thin or optically thick disks
with a  cutoff radius not exceeding 175 AU.  
We obtain a lower limit to the disk mass
from the measured 220 GHz flux by taking the optically thin limit,
$$
M \ge (D^2 F_{\nu}/ \kappa_{\nu} B_{\nu}(T_o)) [2-p-q/2-p] 
(R_D/r_o)^q \eqno(3)
$$
Assuming a cutoff radius $R_D = 175$ AU, an inner
radius  $r_o = 0.01$ AU, an opacity  
$\kappa_\nu = 0.01$ cm$^2$ gm$^{-1}$ at 220 GHz ($\beta=1.5$),
a maximum temperature of $T_{0,max} = 430$ K, and $q \simeq 0.25$  
gives a minimum disk mass  $M \ge 3 \times 10^{-2}$ M$_{\odot}$;
the true disk mass could be much higher.  

\section{Discussion}

The constraints on the size of the central source have
implications for theoretical models  
of the evolution of sources such as VLA 1623.  
We adopt an effective sound speed $a = 0.27 $ km s$^{-1}$ for 
this innermost region of VLA 1623 (\markcite{AWB}Andr\'e et al. 
1993) and a magnetic
field strength $B = 30 ~\mu$G appropriate for low
mass molecular cloud cores (\markcite{myers}Myers \& Goodman 1988). 
In the context of the \markcite{GS}Galli \& Shu (1993) model,
the magnetic support radius of the disk is  given by
$$r_B = 0.12 (G^2B^4/a)^{1/3} t^{7/3} = 
2.75 (B/ 30 \hskip3pt {\rm \mu G})^{4/3} (a/ 0.27 \hskip3pt{\rm km \hskip3pt 
s^{-1}})^{-1/3} t_4^{7/3} \hskip3pt {\rm AU} \eqno(4)$$
where $t_4$ is in units of $10^4$ yrs and $G$ is the gravitational constant.  
If we identify the upper limit to the cutoff radius of 175 AU 
with the radius of the pseudo-disk, $r_B$, then
this model predicts an age for the system of  
$t < 5.9 \times 10^4 (B/ 30 \hskip3pt{\rm \mu G})^{-4/7} (a/ 0.27 
\hskip3pt{\rm km \hskip3pt s^{-1}})^{1/7}$~yr.
This age agrees with general estimates for Class 0 ages 
(\markcite{B94}Barsony 1994), 
but is somewhat longer than initial estimates for the age of
VLA 1623 (\markcite{AWB}Andr\'e et al. 1993).
In a collapsing singular isothermal sphere (SIS) model,
the expansion wave has reached a radius
$r_{exp} = a t_{exp} < 3400 (t_{exp}/5.9 \times 10^4 \hskip3pt
{\rm yr})$ AU, which is comparable to the size of
the 2000 AU envelope seen in the 
\markcite{AWB} Andr\'e et al. (1993) map. In the context of the SIS model,
this age suggests that the envelope and the
collapsed protostellar core+disk will each contain
comparable amounts of material.

Alternatively, if we identify 175 AU with the centrifugal balance
radius, $r_c = 0.058 a \Omega_o^2 t^3 = 
0.030 (\Omega_o / 10^{-13}
s^{-1})^2 t_4^3$ AU, then the age of the system is 
$<1.8\times 10^5$ yr, similar
to age estimates for  Class I sources (\markcite{wly}Wilking,
Lada, \& Young 1989). In this picture, the
magnetic pseudo-disk, if present, would extend to 2300 AU, and must have
 a very low column density in order to be undetectable with
lower resolution measurements
(e.g. \markcite{awb} Andr\'e et al. 1993).
This age is five times longer than the time required for 
the expansion wave 
to reach the edge of the circumstellar envelope. Over
this time, a compressional wave propagating inwards from the envelope
edge would have accelerated the collapse of the remaining 
envelope material onto the stellar core.

Returning to the assumption that most of the compact core is magnetically
supported, 
the centrifugal balance radius in the pseudo-disk 
then occurs at  $r_c < 6.2 (\Omega_o / 10^{-13}
s^{-1})^2$ AU.  This region is  large enough  either to drive off 
a disk wind or to interact with a possible stellar magnetosphere, and
thus theoretical models developed to explain outflows in Class I
sources are likely to be equally valid for Class 0 sources.
The total collapsed mass of the star+disk system in the SIS 
model is $M_{coll} \simeq [1 + (\pi / 4)]^{-1} 2 a^2 r_{exp}/G  = 0.29$ 
M$_{\odot}$. Taking a stellar 
radius $R_* = 2.5$ R$_{\odot}$ 
(valid for models with $M_*< 0.3$ M$_{\odot}$, 
\markcite{stahler}Stahler 1988) 
and a bolometric luminosity $L_* \sim 1$~L$_\odot$ 
(\markcite{AWB}Andr\'e et al. 1993),
we find that an accretion rate through the disk
of $\dot M_a =  L_* R_*/G M_* = 0.27 \times 10^{-6}$ 
M$_{\odot}$~yr$^{-1}$ 
can provide the observed luminosity of
VLA 1623.  This accretion rate is significantly smaller than 
the  rate inferred from the collapse of an SIS model, $\dot M_i =
0.975 a^3/G = 4.5 \times 10^{-6}$ M$_{\odot}$ yr$^{-1}$.  A discrepancy
of a factor of 5 between the disk and collapse accretion rates in SIS
models has been 
noted by \markcite{Kenyon} Kenyon, Calvet, \& Hartmann (1993) 
for more evolved T Tauri systems.   
A possible reason for the inconsistency in the mass 
accretion rates is that SIS models have constant infall rates.   
More general models of molecular cloud cores 
(eg. \markcite{MP} McLaughlin \& Pudritz 1996, 1997, 
\markcite{Foster} Foster \& Chevalier 1993) do not collapse at a constant
accretion rate, which may alleviate this problem.   

Finally, we note that the outflow in VLA 1623 
has reached much larger scales 
($\sim 20000$ AU, \markcite{andre90} Andr\'e et al. 1990)
than the expansion wave predicted by the SIS model.  
This situation is possible because outflows are 
likely to be super-Alfv\'enic.  The effect that such an outflow
has upon collapse has yet to be investigated. 
 
\section{Conclusions} 

High resolution submillimeter continuum observations
have detected a compact source in
VLA 1623, which is most readily interpreted as a disk.
Gaussian fitting gives 
 an upper limit of 70 AU for the half-width half-maximum
radius, which implies an upper limit of 
$\sim 175$ AU for the cutoff radius of the circumstellar disk in the system.  
In the context of existing 
collapse models, such a structure could be
magnetically supported on the largest scales and have an age  
$ \sim 6 \times 10^4$~yr.
The innermost region of the disk within $\sim 6 $ AU is likely to  be in 
centrifugal support, which could provide a drive  
for the outflow according to current theoretical models.   
Alternatively, if the entire observed disk is in centrifugal
support, then the age of the system is $\sim 2\times 10^5$ yr,
similar to age estimates of Class I sources. In this case, we
may be detecting a Class I source viewed nearly edge-on.
 
\acknowledgments

We thank the referee, Lee Mundy, in particular for
pointing our attention to the limit
provided by the  bolometric luminosity of the source.
CDW and REP thank Peter Sutherland for help with numerical
fitting techniques.
The research of REP and CDW is supported through grants
from the Natural Sciences and Engineering Research Council
of Canada.  JEC gratefully acknowledges support from a NSF-YI Award and
a David and Lucile Packard Fellowship.
The JCMT is operated by the Royal Observatories on behalf of the Particle
Physics and Astronomy Research Council of the United Kingdom, the
Netherlands Organization for Scientific Research, and the National Research
Council of Canada. Research at the CSO is supported by NSF grant 90-15755.

\clearpage

\figcaption[figure1.ps]{(a) The visibility flux of VLA 1623 at 220 GHz detected
with the JCMT-CSO interferometer as a function of the projected baseline
length. The flux shows no dependence on projected
baseline length and thus the source is unresolved. 
(b) The visibility flux at 355 GHz as a function of baseline length.
(c) The visibility flux at 355 GHz
as a function of
the hour angle of the observations. 
(d) The source flux at 355 GHz deduced from
50 Monte Carlo simulations as a function of the deduced upper limit to the
source size. With the available data
it is impossible to distinguish between a weak compact source
and a more extended, somewhat stronger source. The dashed line indicates
the 90\% confidence level upper limit to the radius of the disk.
\label{fig-1}}

\figcaption[figure2.ps]{(a)  Cutoff radius
$R_D$ of an optically thick disk in units of the inner
disk radius $r_o$ versus the product $T_o r_o^2$, where $T_o$ is
the temperature of the disk at radius $r_o$. Solid lines are model fits
for different values of $q$, the exponent in the temperature
gradient. 
Viable accretion disk models lie within the region demarcated by
the dashed line segments, A,B,C, described in the text.
The angle between
the disk axis and the line of sight is $\theta = 80^o$.
(b) As in (a), but for optically thin
emission. In this case $R_D$ depends on the product
$\tau_o T_o r_o^2$. 
\label{fig-2}}

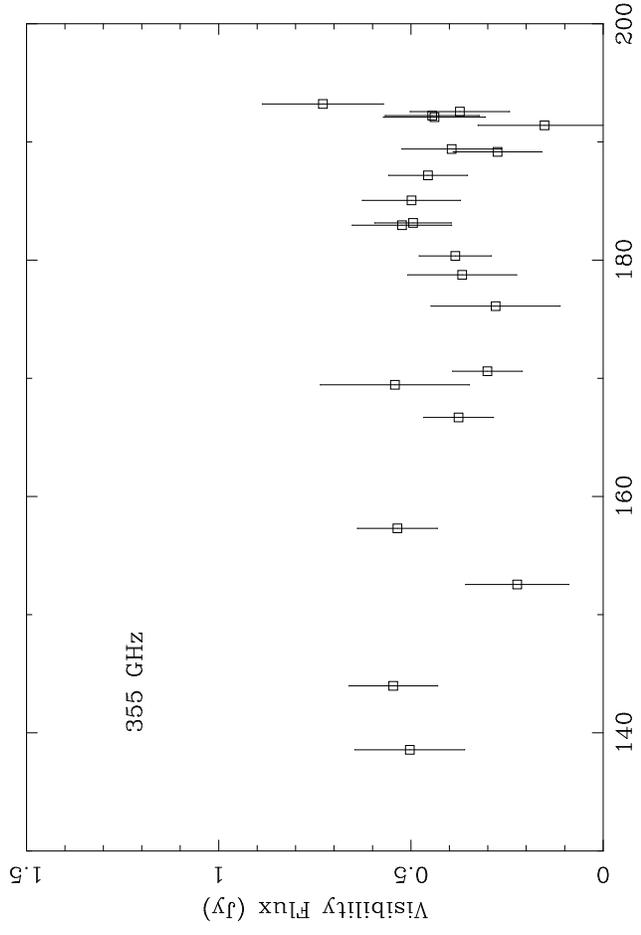
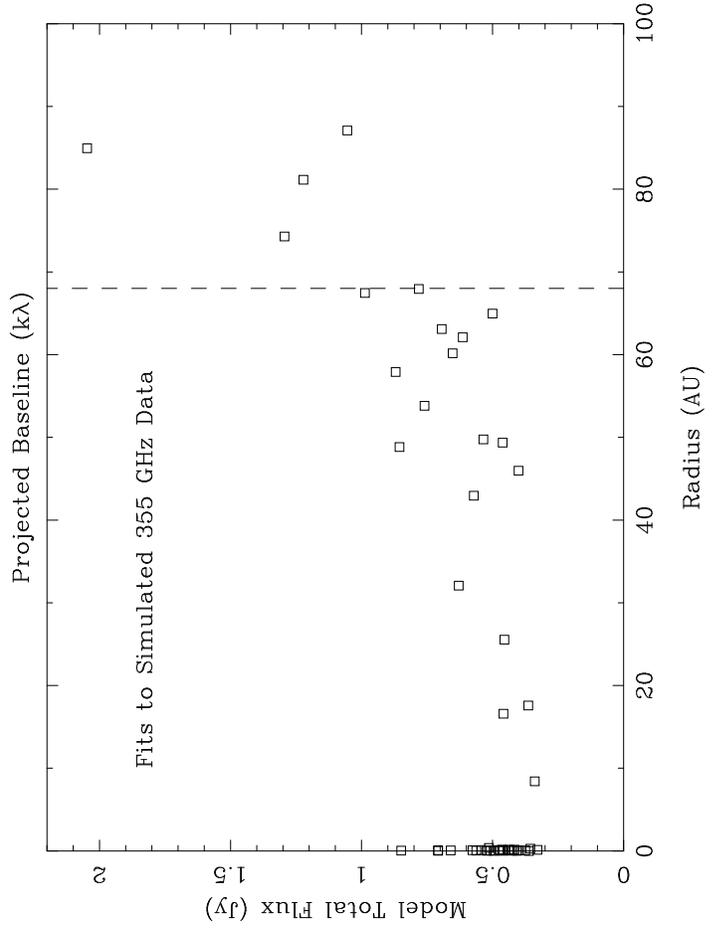
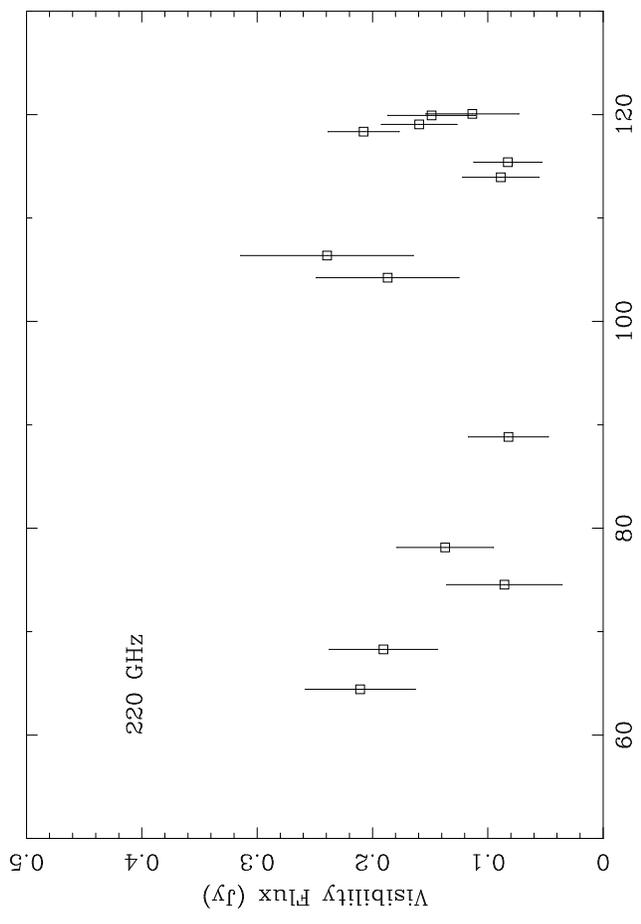
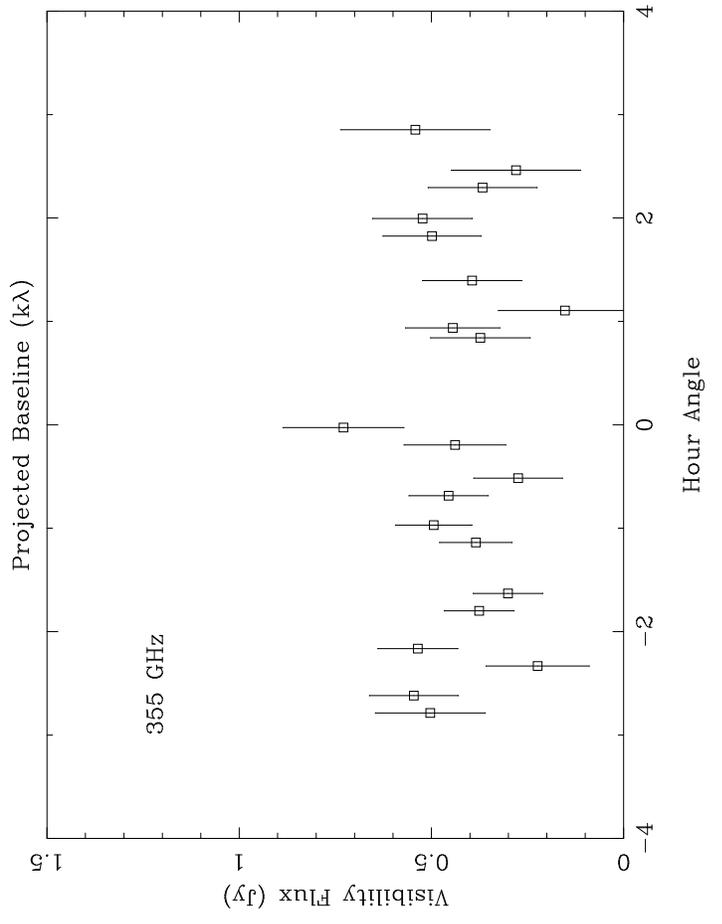
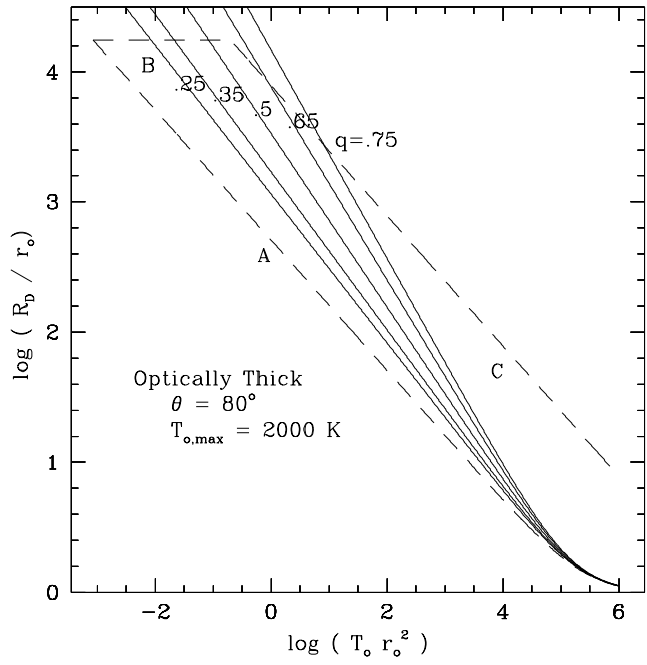
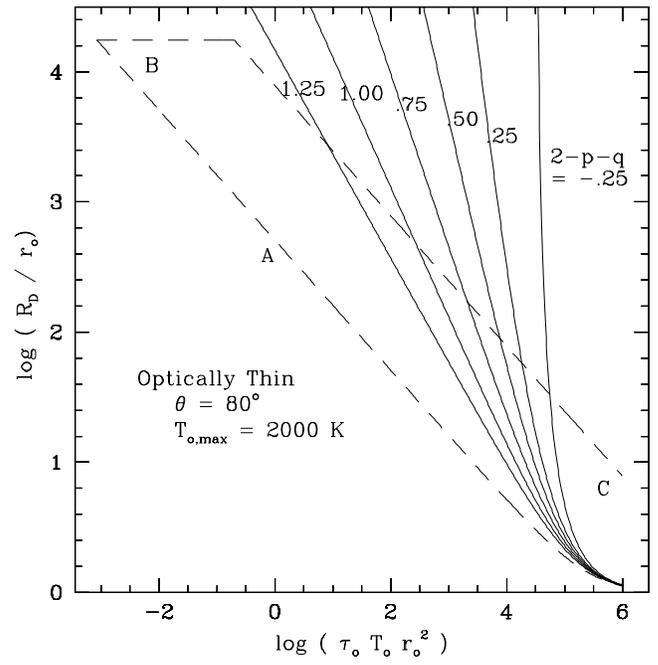
\end{document}